\documentclass[aps,pra,superscriptaddress,reprint,floatfix,longbibliography,showpacs,amsfonts,amsmath,amssymb]{revtex4-1}
\usepackage{graphicx}
\usepackage{hyperref}
\usepackage{xr-hyper}
\usepackage{upgreek}
\usepackage{epstopdf}
\usepackage[note-name=, use-sort-key = false]{notes2bib}
\usepackage{color}
\usepackage{comment}
\usepackage{ulem}

\begin{document}
\title{Pressure-induced excitations in the out-of-plane optical response of the  \\
nodal-line semimetal ZrSiS}

\author{J. Ebad-Allah}
\affiliation{Experimentalphysik II, University of Augsburg, 86159 Augsburg, Germany}
\affiliation{Department of Physics, Tanta University, 31527 Tanta, Egypt}
\author{S. Rojewski}
\affiliation{Experimentalphysik II, University of Augsburg, 86159 Augsburg, Germany}
\author{M. V\"ost}
\affiliation{Chair of Chemical Physics and Materials Science, Institute of Physics, University of Augsburg, 86159 Augsburg,
Germany}
\author{G. Eickerling}
\affiliation{Chair of Chemical Physics and Materials Science, Institute of Physics, University of Augsburg, 86159 Augsburg,
Germany}
\author{W. Scherer}
\affiliation{Chair of Chemical Physics and Materials Science, Institute of Physics, University of Augsburg, 86159 Augsburg,
Germany}
\author{E. Uykur}
\affiliation{1. Physikalisches Institut, Universit\"at Stuttgart, 70569 Stuttgart, Germany}
\author{Raman Sankar}
\affiliation{Institute of Physics, Academia Sinica, Taipei, 11529, Taiwan.}
\author{L. Varrassi}
\affiliation{Department of Physics and Astronomy, Alma Mater Studiorum - Universit\`{a} di Bologna, Bologna, 40127 Italy}
\author{C. Franchini}
\affiliation{University of Vienna, Faculty of Physics and Center for Computational Materials Science, Vienna, Austria}
\affiliation{Department of Physics and Astronomy, Alma Mater Studiorum - Universit\`{a} di Bologna, Bologna, 40127 Italy}
\author{K. Ahn}
\affiliation{Institute of Solid State Physics, TU Wien, 1020 Vienna, Austria}
\author{J. Kune\v{s}}
\affiliation{Institute of Solid State Physics, TU Wien, 1020 Vienna, Austria}
\affiliation{Institute of Physics, The Czech Academy of Sciences, 18221 Praha, Czech Republic}
\author{C. A. Kuntscher}
\email{christine.kuntscher@physik.uni-augsburg.de}
\affiliation{Experimentalphysik II, University of Augsburg, 86159 Augsburg, Germany}

\begin{abstract}
The anisotropic optical response of the layered, nodal-line semimetal ZrSiS at ambient and high pressure is investigated by frequency-dependent reflectivity measurements for the polarization along and perpendicular to the layers. The highly anisotropic optical conductivity is in very good agreement with results from density functional theory calculations and confirms the anisotropic character of ZrSiS. Whereas the in-plane optical conductivity shows only modest pressure-induced changes, we found strong effects on the out-of-plane optical conductivity spectrum of ZrSiS, with the appearance of two prominent excitations.
These pronounced pressure-induced effects can neither be attributed to a structural phase transition according to our single-crystal x-ray diffraction measurements, nor can they be explained by electronic correlation and electron-hole pairing effects, as revealed by theo\-retical calculations. Our findings are discussed in the context of the recently proposed excitonic insulator phase in ZrSiS.
\end{abstract}

\maketitle

%\section{Introduction}

Topological nodal-line semimetals (NLSMs) with linearly dispersing electronic bands, which cross along a line in reciprocal space, host two-dimensional (2D) Dirac fermions and are currently extensively investigated due to their exotic and highly interesting physical properties \cite{Burkov.2011,Fang.2015}. The layered material ZrSiS is considered as the prototype NLSM, where the linearly dispersing bands extend over a large energy range $\sim$2~eV around the Fermi energy E$_F$, without the presence of topologically
trivial bands in the vicinity of E$_F$,  and the corresponding nodal lines form a three-dimensional cage-like structure \cite{Schoop.2016,Neupane.2016,Chen.2017,Pezzini.2018}. There are additional Dirac crossings at the $X$ and $R$ point of the Brillouin zone located $\sim$0.5~eV
above and below E$_F$, which are protected by nonsymmorphic symmetry against gapping due to the spin-orbit coupling.
The unconventional mass enhancement of quasiparticles in ZrSiS \cite{Pezzini.2018} suggests the importance of electronic correlations, which could potentially drive the material toward an excitonic insulator phase or a quantum critical region close to it \cite{Rudenko.2018,Scherer.2018,Wang.2020}.

\begin{figure}[t]
\includegraphics[width=0.35\textwidth]{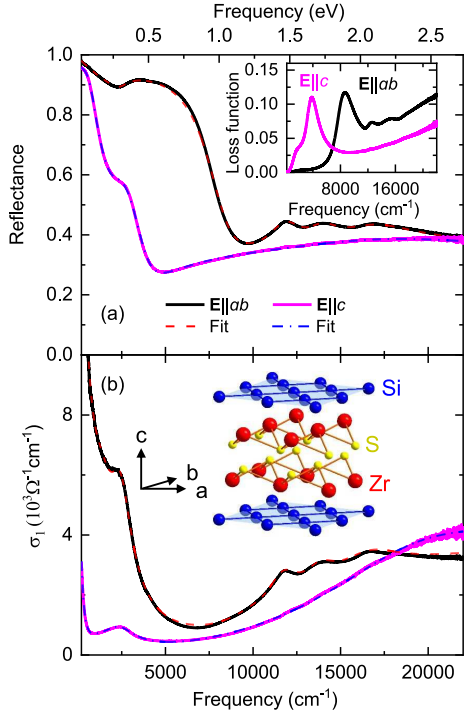}
\caption{Optical response functions of ZrSiS at ambient-conditions
for polarization directions {\bf E}$\parallel$$ab$ and {\bf E}$\parallel$$c$: (a) reflectivity spectra and (b) real part of the optical conductivity, $\sigma_{1}$. Inset of (a): loss function -Im(1/${\hat{\epsilon}}$) where $\hat{\epsilon}$ is the complex dielectric function. Inset of (b): Crystal structure of ZrSiS with Si square nets parallel to the $ab$-plane.}
\label{fig:ambient-pressure}
\end{figure}

The exceptional electronic band structure of ZrSiS and related materials Zr$X$$Y$, where $X$ is a carbon group element ($X$ = Si, Ge, Sn) and $Y$ is a chalcogen element ($Y$ = S, Se, Te) \cite{Wang.1995}, is mainly due
to the 2D square nets of Si atoms parallel to the $a$$b$-plane, which are the main structural motif besides the square nets of Zr and chalcogen atoms, stacked perpendicular to the $a$$b$-plane [see inset of Fig.\ \ref{fig:ambient-pressure}(b)].
Further interesting properties of ZrSiS include high charge carrier mobility and exceptionally large magnetoresistance due to electron-hole
symmetry \cite{Sankar.2017,Lv.2016,Singha.2017}.
Also the electrodynamic properties of ZrSiS are unusual, with a nearly frequency-independent optical conductivity $\sigma_1$ for frequencies from 250 to 2500~cm$^{-1}$ (30-300~meV) \cite{Schilling.2017}. This rather flat behavior of $\sigma_1$ is followed by a U-shaped profile between 3000 and 10.000~cm$^{-1}$ (0.37-1.24~eV), which was ascribed to transitions between the linearly crossing bands of the nodal line close to E$_F$, and a peak located at $\sim$11.800~cm$^{-1}$ ($\sim$1.46~eV), associated with transitions between parallel bands of the Dirac crossings protected by nonsymmorphic symmetry \cite{Ebad-Allah.2019}.

All previous experimental studies on the electrodynamic properties of ZrSiS focused on the in-plane optical response, i.e., for the polarization {\bf E} of the incident electromagnetic radiation aligned along the layers in the $ab$-plane [see inset of Fig.\ \ref{fig:ambient-pressure}(b)]. In this paper, we report on the out-of-plane optical conductivity of ZrSiS, as obtained by frequency-dependent reflectivity measurements for {\bf E} directed perpendicular to the layers, i.e., {\bf E}$\parallel$$c$.
Furthermore, we studied the in-plane and out-of-plane optical conductivity of ZrSiS under external, quasi-hydrostatic pressure, combined with pressure-dependent single crystal x-ray diffraction (XRD) measurements. In the out-of-plane optical response two new excitations appear under pressure, which cannot be reproduced by density functional theory (DFT) calculations at the generalized gradient approximation \cite{Perdew.1996}
level or even by including electronic correlations at GW level and
electron-hole pairing effects. Our findings add yet another interesting facet to the exceptional properties of ZrSiS.

\begin{figure}[t]
\includegraphics[width=0.38\textwidth]{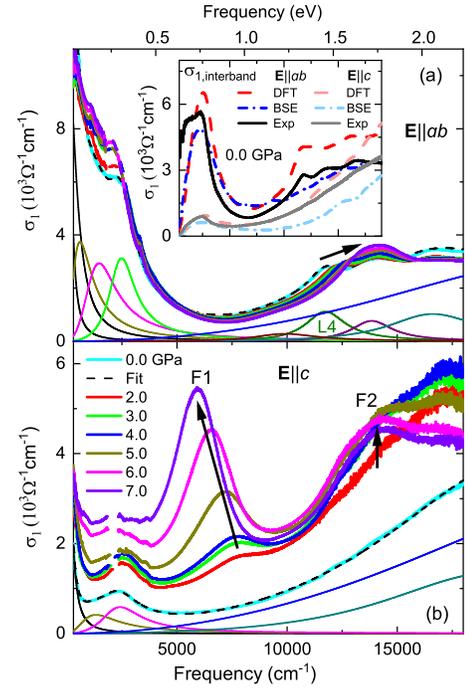}
\caption{Pressure-dependent optical conductivity $\sigma_1$ with the Drude-Lorentz fitting and the corresponding contributions at 0~GPa for (a) {\bf E}$\parallel$$ab$ and (b) {\bf E}$\parallel$$c$. Arrows mark the most pronounced pressure-induced changes in the spectra.
Inset of (a): Comparison between the experimental and both DFT and  BSE calculated interband conductivity $\sigma_{1,interband}$ at 0~GPa.}
\label{fig:conductivity}
\end{figure}

Ambient-pressure reflectivity spectra of ZrSiS for the polarization along ({\bf E}$\parallel$$ab$) and perpendicular ({\bf E}$\parallel$$c$) to the layers are depicted in Fig.\ \ref{fig:ambient-pressure}(a). (See the Supplemental Material \cite{Suppl} for a description of sample preparation, experimental details, and analysis of reflectivity data.)
For both polarization directions, the reflectivity is high at low energies and shows a distinct plasma edge, indicating the metallic character consistent with recent resistivity measurements \cite{Novak.2019,Shirer.2019}. The anisotropic
character of ZrSiS is manifested by the polarization-dependent energy position of the plasma edge, which is shifted towards lower energies for {\bf E}$\parallel$$c$ compared to {\bf E}$\parallel$$ab$. Consistently, the intraband plasmon peak in the loss function defined as -Im(1/${\hat{\epsilon}}$), where $\hat{\epsilon}$ is the complex dielectric function, appears at lower energy, $\approx$0.47~eV, for
{\bf E}$\parallel$$c$ as compared to $\approx$1.07~eV for {\bf E}$\parallel$$ab$ [inset of Fig.\ \ref{fig:ambient-pressure}(a)]. The anisotropic optical response is also seen in the real part of the optical conductivity spectrum $\sigma_1$, displayed in Fig.\ \ref{fig:ambient-pressure}(b). For both directions, $\sigma_1$ consists of a Drude term at low energies due to itinerant charge carriers.
From the spectral weight analysis of the Drude contribution we obtain a plasma frequency $\omega_{p}$ of 3.17 eV for {\bf E}$\parallel$$ab$ and 1.08 eV for {\bf E}$\parallel$$c$, in agreement with the results of first-principles calculations \cite{Zhou.2019}. The ratio of dc conductivities
$\sigma_{ab}/\sigma_{c}$ amounts to $\sim$16, which lies in between the values 8 and 30 given in Refs.\ \cite{Shirer.2019} and \cite{Novak.2019}, respectively. Obviously, the charge dynamics and charge transport is much enhanced along the layers compared to the perpendicular direction, as expected.

Also the profile of the optical conductivity spectrum is strongly polarization-dependent [Fig.\ \ref{fig:ambient-pressure}(b)], in very good agreement with the theoretical results of Refs.\ \cite{Shirer.2019,Habe.2018}. For {\bf E}$\parallel$$ab$, the low-energy $\sigma_1$ spectrum consists of a Drude term and a rather flat region up to $\sim$3000~cm$^{-1}$ followed by a U-shape frequency dependence, which is bounded by a rather sharp peak at high frequencies \cite{Schilling.2017,Ebad-Allah.2019}.
This sharp peak (called L4 in the following) is associated with transitions between parallel bands of the Dirac crossings, which are protected by nonsymmorphic symmetry against gapping \cite{Ebad-Allah.2019}.
The profile of the {\bf E}$\parallel$$c$ optical conductivity is markedly different: It is rather featureless, namely besides the Drude peak it shows only an absorption peak at $\sim$2400~cm$^{-1}$ and a monotonic increase above $\sim$6000~cm$^{-1}$, which originates from transitions between Dirac bands and
states further away from E$_F$.
Compared to {\bf E}$\parallel$$ab$, the out-of plane momentum matrix elements exhibit substantially weaker $k$- and band dependence, and thus the {\bf E}$\parallel$$c$ optical
conductivity reflects the behaviour of the particle-hole (joint) density of states (divided by frequency) \cite{Ebad-Allah.2019}.
For both directions, the optical conductivity and reflectivity spectra can be well fitted with the Drude-Lorentz model (see Fig.\ \ref{fig:ambient-pressure}). The obtained Drude and Lorentz contributions at ambient pressure are shown in Fig.\ \ref{fig:conductivity}(a) and (b) for {\bf E}$\parallel$$ab$ and {\bf E}$\parallel$$c$, resp..
For comparison with the results of DFT calculations (see the Suppl.\ Material \cite{Suppl} for details)
we subtracted the Drude term from the total $\sigma_1$ spectrum and obtained the contributions from the interband transitions, $\sigma_{1,interband}$.
The interband conductivity spectra agree well with the corresponding theoretical spectra
[see inset of Fig.\ \ref{fig:conductivity}(a)].

\begin{figure*}
\includegraphics[width=1\textwidth]{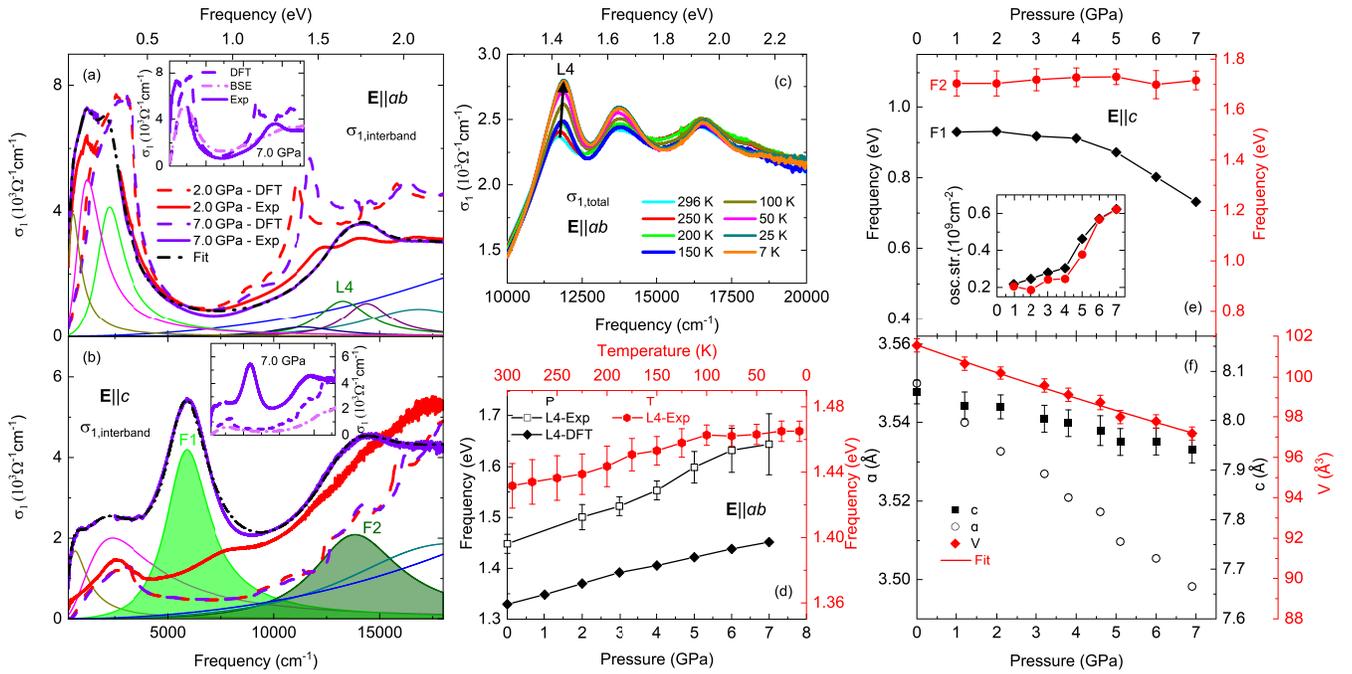}
\caption{(a),(b) Comparison between the experimental and DFT interband conductivity $\sigma_{1,interband}$ at 2.0 and 7.0~GPa, with the total fitting curve and the fitting contributions for the experimental $\sigma_{1,interband}$ at 7.0~GPa, for {\bf E}$\parallel$$ab$ and {\bf E}$\parallel$$c$, respectively. Insets of (a),(b): Comparison between DFT and BSE theoretical results and experimental interband conductivity $\sigma_1$ at 7.0 GPa for {\bf E}$\parallel$$ab$ and {\bf E}$\parallel$$c$, respectively. (c) Temperature-dependent high-energy optical conductivity for {\bf E}$\parallel$$ab$. The arrow marks the temperature-induced shift of the L4 peak. (d) Comparison between the temperature ($T$) and pressure ($P$) dependence (both experimental and theoretical) of the frequency position of the L4 peak. (e) Pressure-dependent frequency position and oscillator strength (inset) of the peaks F1 and F2. (f) Volume $V$ of the unit cell and lattice parameters $a$ and $b$ as a function of pressure. The solid line is a fit with a Murnaghan-type EOS (see text).}
\label{fig:Figure3}
\end{figure*}

In the following, the main focus will be on the optical response of ZrSiS under external pressure. The experimental in-plane and out-of-plane $\sigma_1$ spectra are depicted for selected pressures in Figs.\ \ref{fig:conductivity}(a) and (b), respectively. First, we
discuss the results for the in-plane optical response. One notices that the induced changes for {\bf E}$\parallel$$ab$ are only modest, and the characteristic profile of the optical conductivity with its U-shape is unchanged up to 7~GPa. A detailed analysis reveals a slight increase of $\sigma_1$ below $\sim$3000~cm$^{-1}$ and a shift of the L4 peak to higher energies with increasing pressure. A comparison between the experimental and theoretical interband optical conductivity from DFT calculations for {\bf E}$\parallel$$ab$ is given in Fig.\ \ref{fig:Figure3}(a) for two selected pressures (2.0 and 7.0~GPa). Like for the experimental results, the U-shape of the theoretical spectrum
is retained up to the highest measured pressure and the L4 peak at the high-energy bound of the U-shape is blue-shifted. According to the behavior of the L4 peak, pressure induces a shift of the nonsymmorphic-symmetry protected Dirac crossings
{\it away} from E$_F$, as a result of the compression of the crystal lattice.

Consistently, the thermal contraction of the crystal lattice during cooling causes a blue shift of the L4 peak [see  Fig.\ \ref{fig:Figure3}(c)] \bibnote{Please note that we concentrate here on the high-energy range, since the temperature dependence of the low-energy optical conductivity for {\bf E}$\parallel$$ab$ has been discussed in detail in Ref.\ \cite{Schilling.2017}.}.
A comparison between the effect of cooling and pressure on the energy position of the L4 peak is given in Fig.\ \ref{fig:Figure3}(d), whereby for the latter
both experimental and DFT results are displayed
\bibnote{It is interesting to note, that the temperature-dependent shift of the L4 peak slightly changes its slope between 150 and 100~K. In the same temperature range, the temperature dependence of the dc resistivity changes \cite{Singha.2017} and several Raman modes show an anomaly in their position and width \cite{Singha.2018}, which was attributed to an interplay between electron and phonon degrees of freedom.}.
To conclude, tensile strain, instead of compressive strain, would be needed to push the non-symmorphic symmetry protected Dirac nodes in ZrSiS {\it toward} E$_F$, in order to study the expected distinct physics related to these 2D Dirac fermions \cite{Young.2015}.

Next, we will discuss the pressure-induced effects on the out-of-plane optical conductivity [see Fig.\ \ref{fig:conductivity}(b)].
Starting from the lowest applied pressure (2~GPa), drastic changes occur in the profile of the {\bf E}$\parallel$$c$ $\sigma_1$ spectrum: besides the pressure-induced increase below $\sim$5000~cm$^{-1}$,
two new excitations labelled F1 and F2 appear, which gain spectral weight with increasing pressure \bibnote{We cannot completely rule out that the F2 peak already exists at ambient pressure, but cannot be resolved due to small oscillator strength and overlap with higher-energy interband transitions.}.
Similar to the in-plane optical response, a Drude-Lorentz model was applied for fitting the experimental spectra.
As an example, we depict in Fig.\ \ref{fig:Figure3}(b) the experimental interband conductivity $\sigma_{1,interband}$ at 7~GPa, where the Drude term was subtracted from the total $\sigma_1$, together with the Lorentz contributions.
Each of the two new excitations F1 and F2 can be well described by one Lorentzian term. With increasing pressure, the energy position of excitation F2 is almost unchanged, whereas F1 first shifts slightly to lower energies for pressures up to $\sim$4~GPa, and for pressures above 4~GPa this redshift gets more pronounced [Fig.\ \ref{fig:Figure3}(e)]. The oscillator strength of F1 and F2 slightly increases with increasing pressure up to 4~GPa and increases strongly above $\sim$4~GPa [inset of Fig.\ \ref{fig:Figure3}(e)]
due to a transfer of spectral weight from the energy range above $\sim$1.9~eV.
In Fig.\ \ref{fig:Figure3}(b) we compare the experimental {\bf E}$\parallel$$c$ $\sigma_{1,interband}$ spectrum for two selected pressures 2.0 and 7.0~GPa with the corresponding DFT results. Interestingly, the theoretical interband conductivity for {\bf E}$\parallel$$c$ is basically unchanged upon pressure application, in strong contrast to the experimental results. In particular, the two excitations F1 and F2 are {\it not} reproduced in the pressure-dependent theoretical spectra.
Thus, the excitations F1 and F2 should be attributed to effects, which are not included in the band structure calculations, and hence the role of beyond-DFT effects might be relevant, as discussed below.

For an interpretation of our findings, we performed a high-pressure XRD study on a ZrSiS single crystal (see the Suppl.\ Material \cite{Suppl} for details). With increasing pressure, the lattice parameters $a$ and $c$ monotonically decrease, resulting in a monotonic volume decrease [see Fig.\ \ref{fig:Figure3}(f)]. Further investigation of the reciprocal space up to maximum measured pressure (6.9 GPa) does not reveal the formation of additional or superstructural Bragg reflections \cite{Suppl}. Hence, our diffraction data do not provide any hint for a structural phase transition up to at least $\sim$7~GPa, in agreement with Refs.\ \cite{Gu.2019,VanGennep.2019}. This finding is in contradiction with results of powder XRD experiments, which suggested a structural phase transition at elevated pressures \cite{Singha.2018}.
We fitted the volume $V$ with the second-order Murnaghan equation of state (EOS) \cite{Murnaghan.1944} according to $V(p)=V_{0}\cdot\lbrack(B_{0}^{\prime}/B_{0})\cdot p+1]^{-1/{B_{0}^{\prime}}}$, where $B_{0}$ is the bulk modulus, $B_{0}^{\prime}$ its pressure derivative, which is fixed to 4, and $V_0$ the volume, all at $P$=0~GPa. From the fitting we obtain the value $B_{0}$=144$\pm 5$~GPa, consistent with earlier reports  \cite{Singha.2018,Salmankurt.2016}.

Based on the results of our high-pressure XRD study, we can discuss the optical data in more detail.
First, it has been proposed that the distance of the nonsymmorphic Dirac crossings from E$_F$ is inversely proportional to the distance between the Si atoms in the Si-Si square nets \cite{Kirby.2020} and hence inversely proportional to the lattice parameter $a$.
Accordingly, the energy position of the related L4 peak in the {\bf E}$\parallel$$ab$ $\sigma_1$ spectrum should scale with 1/$a$.
However, based on our pressure-dependent optical data, we cannot confirm such a behavior.
Second, the experimental optical data show an anomaly at $\approx$4~GPa in the shift of the F1 excitation and in the oscillator strength of the F1 and F2 excitations. Since our pressure-dependent XRD results do not provide any evidence for a structural phase transition, this anomaly arises from purly electronic behavior, like in pressurized ZrSiTe \cite{Ebad-Allah.2019a,Krottenmuller.2020}.
The origin of the excitations F1 and F2, which appear under pressure in the experimental {\bf E}$\parallel$$c$ optical conductivity remains, however, unclear.

It is interesting to note that, based on calculations for a
bi-layer square lattice model, Rudenko {\it et al.} \cite{Rudenko.2018}
suggested that ZrSiS undergoes a condensation of inter-layer zero-momentum excitons due to electronic correlations and a high degree of electron-hole symmetry of the electronic band structure, which gives rise to an excitonic insulator state at low temperature.
In this weak-coupling scenario (formally similar to BCS superconductivity), a gap
opens at E$_F$ in the excitonic insulator state, which leads to a spectral weight
transfer and appearance of Hebel-Slichter-like peaks~\cite{Hebel1959,Rudenko.2018}. Transitions between these peaks could in principle lead to distinct excitations in the conductivity spectrum
\bibnote{We note that the F1 peak derives its spectral weight from the high-energy rather than low-energy region. This is opposite to the excitonic insulator scenario, where the spectral weight comes from the Drude peak or low-energy region in general.}.
However, the signatures of the exciton instability have not been experimentally observed in ZrSiS until now, in particular no pseudogap was observed in photoemission spectra.
%The insensitivity of $\sigma_{ab}$ to pressure as well as the fact %that the F1 peak feeds from high- rather than low-energy spectral %weight contradict this scenario.

To inspect the role of zero-momentum excitons
in the formation of the F1 and F2 peaks we have computed the interband optical spectra by solving the Bethe-Salpeter equation (BSE) with quasiparticle energies calculated at GW level, also testing the impact of the coupling between the resonant and anti-resonant excitations.
The results, depicted in the insets of Fig.~\ref{fig:conductivity}(a) and  Figs.~\ref{fig:Figure3}(a) and (b),
show that the optical conductivity is only marginally affected by the electronic correlation and excitonic effects: the BSE $\sigma_1$ spectra are very similar to the DFT spectra and do not exhibit any evident pressure dependence.
Formation of a single exciton thus cannot explain our experimental results. A more complex possibility would be the creation of a finite-momentum exciton accompanied by a phonon or another exciton, in order to ensure the momentum conservation. A first step in analysis of such a scenario would be extending the BSE analysis to finite momentum transfer.

Another scenario was proposed recently \cite{Wang.2020}
suggesting that ZrSiS should be located in a quantum critical region between the NLSM and excitonic insulator phases, which could explain the observed quasiparticle mass enhancement  \cite{Pezzini.2018} in the absence of a pseudogap, consistent with reported photoemission spectra and our optical data. Nevertheless, both the excitonic insulator and quantum critical scenarios are at odds with our observation of pressure-insensitive in-plane response $\sigma_{ab}$ and with our theoretical predications.
We note that a purely electronic excitonic insulator phase with permanent out-of-plane electric dipole moments arranged in an antiferroelectric pattern was recently proposed in bulk MoS$_2$ under pressure \cite{Samaneh.2021}, which might be relevant for pressurized ZrSiS as well.

In conclusion, according to our reflectivity study, the optical response of the NLSM ZrSiS is highly anisotropic. The polarization-dependent optical conductivity at ambient pressure is in very good agreement with the results of DFT calculations. The in-plane optical response shows only modest changes under pressure, consistent with theoretical predictions. In stark contrast, the out-of-plane optical conductivity undergoes strong changes under pressure, with the appearance of two pronounced peaks.
%With increasing pressure the low-energy peak F1 softens %and %its %spectral weight increases fed by transfer from %higher %energies.
The observed pressure-induced changes can neither
be attributed to a structural phase transition according to our single-crystal XRD data, nor can they be explained by electronic correlation effects and single exciton formation according to our theoretical calculations.

\begin{acknowledgments}
  We thank S. Sharma for fruitful discussions. C.A.K. acknowledges
  financial support from the Deutsche Forschungsgemeinschaft (DFG),
  Germany, through grant no.\ KU 1432/13-1.  RS acknowledges financial support provided by the Ministry of Science and Technology in Taiwan under project number MOST-108-2112-M-001-049-MY2 and acknowledge for the Academia Sinica for the budget of AS-iMATE-109-13.
\end{acknowledgments}


\begin{thebibliography}{56}%
\makeatletter
\providecommand \@ifxundefined [1]{%
 \@ifx{#1\undefined}
}%
\providecommand \@ifnum [1]{%
 \ifnum #1\expandafter \@firstoftwo
 \else \expandafter \@secondoftwo
 \fi
}%
\providecommand \@ifx [1]{%
 \ifx #1\expandafter \@firstoftwo
 \else \expandafter \@secondoftwo
 \fi
}%
\providecommand \natexlab [1]{#1}%
\providecommand \enquote  [1]{``#1''}%
\providecommand \bibnamefont  [1]{#1}%
\providecommand \bibfnamefont [1]{#1}%
\providecommand \citenamefont [1]{#1}%
\providecommand \href@noop [0]{\@secondoftwo}%
\providecommand \href [0]{\begingroup \@sanitize@url \@href}%
\providecommand \@href[1]{\@@startlink{#1}\@@href}%
\providecommand \@@href[1]{\endgroup#1\@@endlink}%
\providecommand \@sanitize@url [0]{\catcode `\\12\catcode `\$12\catcode
  `\&12\catcode `\#12\catcode `\^12\catcode `\_12\catcode `\%12\relax}%
\providecommand \@@startlink[1]{}%
\providecommand \@@endlink[0]{}%
\providecommand \url  [0]{\begingroup\@sanitize@url \@url }%
\providecommand \@url [1]{\endgroup\@href {#1}{\urlprefix }}%
\providecommand \urlprefix  [0]{URL }%
\providecommand \Eprint [0]{\href }%
\providecommand \doibase [0]{http://dx.doi.org/}%
\providecommand \selectlanguage [0]{\@gobble}%
\providecommand \bibinfo  [0]{\@secondoftwo}%
\providecommand \bibfield  [0]{\@secondoftwo}%
\providecommand \translation [1]{[#1]}%
\providecommand \BibitemOpen [0]{}%
\providecommand \bibitemStop [0]{}%
\providecommand \bibitemNoStop [0]{.\EOS\space}%
\providecommand \EOS [0]{\spacefactor3000\relax}%
\providecommand \BibitemShut  [1]{\csname bibitem#1\endcsname}%
\let\auto@bib@innerbib\@empty
%</preamble>
\bibitem [{\citenamefont {Burkov}\ and\ \citenamefont
  {Balents}(2011)}]{Burkov.2011}%
  \BibitemOpen
  \bibfield  {author} {\bibinfo {author} {\bibfnamefont {A.~A.}\ \bibnamefont
  {Burkov}}\ and\ \bibinfo {author} {\bibfnamefont {L.}~\bibnamefont
  {Balents}},\ }\href@noop {} {\bibfield  {journal} {\bibinfo  {journal} {Phys.
  Rev. Lett.}\ }\textbf {\bibinfo {volume} {107}},\ \bibinfo {pages} {127205}
  (\bibinfo {year} {2011})}\BibitemShut {NoStop}%
\bibitem [{\citenamefont {Fang}\ \emph {et~al.}(2015)\citenamefont {Fang},
  \citenamefont {Chen}, \citenamefont {Kee},\ and\ \citenamefont
  {Fu}}]{Fang.2015}%
  \BibitemOpen
  \bibfield  {author} {\bibinfo {author} {\bibfnamefont {C.}~\bibnamefont
  {Fang}}, \bibinfo {author} {\bibfnamefont {Y.}~\bibnamefont {Chen}}, \bibinfo
  {author} {\bibfnamefont {H.-Y.}\ \bibnamefont {Kee}}, \ and\ \bibinfo
  {author} {\bibfnamefont {L.}~\bibnamefont {Fu}},\ }\href@noop {} {\bibfield
  {journal} {\bibinfo  {journal} {Phys. Rev. B}\ }\textbf {\bibinfo {volume}
  {92}},\ \bibinfo {pages} {081201(R)} (\bibinfo {year} {2015})}\BibitemShut
  {NoStop}%
\bibitem [{\citenamefont {Schoop}\ \emph {et~al.}(2016)\citenamefont {Schoop},
  \citenamefont {Ali}, \citenamefont {Strasser}, \citenamefont {Duppel},
  \citenamefont {Parkin}, \citenamefont {Lotsch},\ and\ \citenamefont
  {Ast}}]{Schoop.2016}%
  \BibitemOpen
  \bibfield  {author} {\bibinfo {author} {\bibfnamefont {L.~M.}\ \bibnamefont
  {Schoop}}, \bibinfo {author} {\bibfnamefont {M.~N.}\ \bibnamefont {Ali}},
  \bibinfo {author} {\bibfnamefont {C.}~\bibnamefont {Strasser}}, \bibinfo
  {author} {\bibfnamefont {V.}~\bibnamefont {Duppel}}, \bibinfo {author}
  {\bibfnamefont {S.~S.~P.}\ \bibnamefont {Parkin}}, \bibinfo {author}
  {\bibfnamefont {B.~V.}\ \bibnamefont {Lotsch}}, \ and\ \bibinfo {author}
  {\bibfnamefont {C.~R.}\ \bibnamefont {Ast}},\ }\bibfield  {title} {\enquote
  {\bibinfo {title} {Dirac cone protected by non-symmorphic symmetry and
  three-dimensional dirac line node in {ZrSiS}},}\ }\href@noop {} {\bibfield
  {journal} {\bibinfo  {journal} {Nat. Commun.}\ }\textbf {\bibinfo {volume}
  {7}},\ \bibinfo {pages} {11696} (\bibinfo {year} {2016})}\BibitemShut
  {NoStop}%
\bibitem [{\citenamefont {Neupane}\ \emph {et~al.}(2016)\citenamefont
  {Neupane}, \citenamefont {Belopolski}, \citenamefont {Hosen}, \citenamefont
  {Sanchez}, \citenamefont {Sankar}, \citenamefont {Szlawska}, \citenamefont
  {Xu}, \citenamefont {Dimitri}, \citenamefont {Dhakal}, \citenamefont
  {Maldonado}, \citenamefont {Oppeneer}, \citenamefont {Kaczorowski},
  \citenamefont {Chou}, \citenamefont {Hasan},\ and\ \citenamefont
  {Durakiewicz}}]{Neupane.2016}%
  \BibitemOpen
  \bibfield  {author} {\bibinfo {author} {\bibfnamefont {M.}~\bibnamefont
  {Neupane}}, \bibinfo {author} {\bibfnamefont {I.}~\bibnamefont {Belopolski}},
  \bibinfo {author} {\bibfnamefont {M.~M.}\ \bibnamefont {Hosen}}, \bibinfo
  {author} {\bibfnamefont {D.~S.}\ \bibnamefont {Sanchez}}, \bibinfo {author}
  {\bibfnamefont {R.}~\bibnamefont {Sankar}}, \bibinfo {author} {\bibfnamefont
  {M.}~\bibnamefont {Szlawska}}, \bibinfo {author} {\bibfnamefont {S.-Y.}\
  \bibnamefont {Xu}}, \bibinfo {author} {\bibfnamefont {K.}~\bibnamefont
  {Dimitri}}, \bibinfo {author} {\bibfnamefont {N.}~\bibnamefont {Dhakal}},
  \bibinfo {author} {\bibfnamefont {P.}~\bibnamefont {Maldonado}}, \bibinfo
  {author} {\bibfnamefont {P.~M.}\ \bibnamefont {Oppeneer}}, \bibinfo {author}
  {\bibfnamefont {D.}~\bibnamefont {Kaczorowski}}, \bibinfo {author}
  {\bibfnamefont {F.}~\bibnamefont {Chou}}, \bibinfo {author} {\bibfnamefont
  {M.~Z.}\ \bibnamefont {Hasan}}, \ and\ \bibinfo {author} {\bibfnamefont
  {T.}~\bibnamefont {Durakiewicz}},\ }\bibfield  {title} {\enquote {\bibinfo
  {title} {Observation of topological nodal fermion semimetal phase in
  {ZrSiS}},}\ }\href@noop {} {\bibfield  {journal} {\bibinfo  {journal} {Phys.
  Rev. B}\ }\textbf {\bibinfo {volume} {93}},\ \bibinfo {pages} {201104(R)}
  (\bibinfo {year} {2016})}\BibitemShut {NoStop}%
\bibitem [{\citenamefont {Chen}\ \emph {et~al.}(2017)\citenamefont {Chen},
  \citenamefont {Xu}, \citenamefont {Jiang}, \citenamefont {Wu}, \citenamefont
  {Qi}, \citenamefont {Yang}, \citenamefont {Wang}, \citenamefont {Sun},
  \citenamefont {Schr\"oter}, \citenamefont {Yang}, \citenamefont {Schoop},
  \citenamefont {Lv}, \citenamefont {Zhou}, \citenamefont {Chen}, \citenamefont
  {Yao}, \citenamefont {Lu}, \citenamefont {Chen}, \citenamefont {Felser},
  \citenamefont {Yan}, \citenamefont {Liu},\ and\ \citenamefont
  {Chen}}]{Chen.2017}%
  \BibitemOpen
  \bibfield  {author} {\bibinfo {author} {\bibfnamefont {C.}~\bibnamefont
  {Chen}}, \bibinfo {author} {\bibfnamefont {X.}~\bibnamefont {Xu}}, \bibinfo
  {author} {\bibfnamefont {J.}~\bibnamefont {Jiang}}, \bibinfo {author}
  {\bibfnamefont {S.-C.}\ \bibnamefont {Wu}}, \bibinfo {author} {\bibfnamefont
  {Y.~P.}\ \bibnamefont {Qi}}, \bibinfo {author} {\bibfnamefont {L.~X.}\
  \bibnamefont {Yang}}, \bibinfo {author} {\bibfnamefont {M.~X.}\ \bibnamefont
  {Wang}}, \bibinfo {author} {\bibfnamefont {Y.}~\bibnamefont {Sun}}, \bibinfo
  {author} {\bibfnamefont {N.~B.~M.}\ \bibnamefont {Schr\"oter}}, \bibinfo
  {author} {\bibfnamefont {H.~F.}\ \bibnamefont {Yang}}, \bibinfo {author}
  {\bibfnamefont {L.~M.}\ \bibnamefont {Schoop}}, \bibinfo {author}
  {\bibfnamefont {Y.~Y.}\ \bibnamefont {Lv}}, \bibinfo {author} {\bibfnamefont
  {J.}~\bibnamefont {Zhou}}, \bibinfo {author} {\bibfnamefont {Y.~B.}\
  \bibnamefont {Chen}}, \bibinfo {author} {\bibfnamefont {S.~H.}\ \bibnamefont
  {Yao}}, \bibinfo {author} {\bibfnamefont {M.~H.}\ \bibnamefont {Lu}},
  \bibinfo {author} {\bibfnamefont {Y.~F.}\ \bibnamefont {Chen}}, \bibinfo
  {author} {\bibfnamefont {C.}~\bibnamefont {Felser}}, \bibinfo {author}
  {\bibfnamefont {B.~H.}\ \bibnamefont {Yan}}, \bibinfo {author} {\bibfnamefont
  {Z.~K.}\ \bibnamefont {Liu}}, \ and\ \bibinfo {author} {\bibfnamefont
  {Y.~L.}\ \bibnamefont {Chen}},\ }\bibfield  {title} {\enquote {\bibinfo
  {title} {Dirac line nodes and effect of spin-orbit coupling in the
  nonsymmorphic critical semimetals {$M$SiS ($M$=Hf, Zr)}},}\ }\href@noop {}
  {\bibfield  {journal} {\bibinfo  {journal} {Phys. Rev. B}\ }\textbf {\bibinfo
  {volume} {95}},\ \bibinfo {pages} {125126} (\bibinfo {year}
  {2017})}\BibitemShut {NoStop}%
\bibitem [{\citenamefont {Pezzini}\ \emph {et~al.}(2018)\citenamefont
  {Pezzini}, \citenamefont {van Delft}, \citenamefont {Schoop}, \citenamefont
  {Lotsch}, \citenamefont {Carrington}, \citenamefont {Katsnelson},
  \citenamefont {Husse},\ and\ \citenamefont {Wiedmann}}]{Pezzini.2018}%
  \BibitemOpen
  \bibfield  {author} {\bibinfo {author} {\bibfnamefont {S.}~\bibnamefont
  {Pezzini}}, \bibinfo {author} {\bibfnamefont {M.~R.}\ \bibnamefont {van
  Delft}}, \bibinfo {author} {\bibfnamefont {L.}~\bibnamefont {Schoop}},
  \bibinfo {author} {\bibfnamefont {B.}~\bibnamefont {Lotsch}}, \bibinfo
  {author} {\bibfnamefont {A.}~\bibnamefont {Carrington}}, \bibinfo {author}
  {\bibfnamefont {M.~I.}\ \bibnamefont {Katsnelson}}, \bibinfo {author}
  {\bibfnamefont {N.~E.}\ \bibnamefont {Husse}}, \ and\ \bibinfo {author}
  {\bibfnamefont {S.}~\bibnamefont {Wiedmann}},\ }\bibfield  {title} {\enquote
  {\bibinfo {title} {Unconventional mass enhancement around the dirac nodal
  loop in {ZrSiS}},}\ }\href@noop {} {\bibfield  {journal} {\bibinfo  {journal}
  {Nat. Phys.}\ }\textbf {\bibinfo {volume} {14}},\ \bibinfo {pages} {178}
  (\bibinfo {year} {2018})}\BibitemShut {NoStop}%
\bibitem [{\citenamefont {Rudenko}\ \emph {et~al.}(2018)\citenamefont
  {Rudenko}, \citenamefont {Stepanov}, \citenamefont {Lichtenstein},\ and\
  \citenamefont {Katsnelson}}]{Rudenko.2018}%
  \BibitemOpen
  \bibfield  {author} {\bibinfo {author} {\bibfnamefont {A.~N.}\ \bibnamefont
  {Rudenko}}, \bibinfo {author} {\bibfnamefont {E.~A.}\ \bibnamefont
  {Stepanov}}, \bibinfo {author} {\bibfnamefont {A.~I.}\ \bibnamefont
  {Lichtenstein}}, \ and\ \bibinfo {author} {\bibfnamefont {M.~I.}\
  \bibnamefont {Katsnelson}},\ }\bibfield  {title} {\enquote {\bibinfo {title}
  {Excitonic instability and pseudogap formation in nodal line semimetal
  {ZrSiS}},}\ }\href@noop {} {\bibfield  {journal} {\bibinfo  {journal} {Phys.
  Rev. Lett.}\ }\textbf {\bibinfo {volume} {120}},\ \bibinfo {pages} {216401}
  (\bibinfo {year} {2018})}\BibitemShut {NoStop}%
\bibitem [{\citenamefont {Scherer}\ \emph {et~al.}(2018)\citenamefont
  {Scherer}, \citenamefont {Honerkamp}, \citenamefont {Rudenko}, \citenamefont
  {Stepanov}, \citenamefont {Lichtenstein},\ and\ \citenamefont
  {Katsnelson}}]{Scherer.2018}%
  \BibitemOpen
  \bibfield  {author} {\bibinfo {author} {\bibfnamefont {M.~M.}\ \bibnamefont
  {Scherer}}, \bibinfo {author} {\bibfnamefont {C.}~\bibnamefont {Honerkamp}},
  \bibinfo {author} {\bibfnamefont {A.~N.}\ \bibnamefont {Rudenko}}, \bibinfo
  {author} {\bibfnamefont {E.~A.}\ \bibnamefont {Stepanov}}, \bibinfo {author}
  {\bibfnamefont {A.~I.}\ \bibnamefont {Lichtenstein}}, \ and\ \bibinfo
  {author} {\bibfnamefont {M.~I.}\ \bibnamefont {Katsnelson}},\ }\bibfield
  {title} {\enquote {\bibinfo {title} {Excitonic insitability and
  unconventional pairing in the nodal-line materials {ZrSiS} and {ZrSiSe}},}\
  }\href@noop {} {\bibfield  {journal} {\bibinfo  {journal} {Phys. Rev. B}\
  }\textbf {\bibinfo {volume} {98}},\ \bibinfo {pages} {241112(R)} (\bibinfo
  {year} {2018})}\BibitemShut {NoStop}%
\bibitem [{\citenamefont {Wang}\ \emph {et~al.}(2020)\citenamefont {Wang},
  \citenamefont {Liu}, \citenamefont {Wan},\ and\ \citenamefont
  {Zhang}}]{Wang.2020}%
  \BibitemOpen
  \bibfield  {author} {\bibinfo {author} {\bibfnamefont {J.-R.}\ \bibnamefont
  {Wang}}, \bibinfo {author} {\bibfnamefont {G.-Z.}\ \bibnamefont {Liu}},
  \bibinfo {author} {\bibfnamefont {X.}~\bibnamefont {Wan}}, \ and\ \bibinfo
  {author} {\bibfnamefont {C.}~\bibnamefont {Zhang}},\ }\bibfield  {title}
  {\enquote {\bibinfo {title} {Quantum criticality of the excitonic insulating
  transition in the nodal-line semimetal {ZrSiS}},}\ }\href@noop {} {\bibfield
  {journal} {\bibinfo  {journal} {Phys. Rev. B}\ }\textbf {\bibinfo {volume}
  {101}},\ \bibinfo {pages} {245151} (\bibinfo {year} {2020})}\BibitemShut
  {NoStop}%
\bibitem [{\citenamefont {Wang}\ and\ \citenamefont
  {Hughbanks}(1995)}]{Wang.1995}%
  \BibitemOpen
  \bibfield  {author} {\bibinfo {author} {\bibfnamefont {C.}~\bibnamefont
  {Wang}}\ and\ \bibinfo {author} {\bibfnamefont {T.}~\bibnamefont
  {Hughbanks}},\ }\bibfield  {title} {\enquote {\bibinfo {title} {Main group
  element size and substitution effects on the structural dimensionality of
  zirconium tellurides of the {ZrSiS} type},}\ }\href@noop {} {\bibfield
  {journal} {\bibinfo  {journal} {Inorg. Chem.}\ }\textbf {\bibinfo {volume}
  {34}},\ \bibinfo {pages} {5524} (\bibinfo {year} {1995})}\BibitemShut
  {NoStop}%
\bibitem [{\citenamefont {Sankar}\ \emph {et~al.}(2017)\citenamefont {Sankar},
  \citenamefont {Peramaiyan}, \citenamefont {Muthuselvam}, \citenamefont
  {Butler}, \citenamefont {Dimitri}, \citenamefont {Neupane}, \citenamefont
  {Rao}, \citenamefont {Lin},\ and\ \citenamefont {Chou}}]{Sankar.2017}%
  \BibitemOpen
  \bibfield  {author} {\bibinfo {author} {\bibfnamefont {R.}~\bibnamefont
  {Sankar}}, \bibinfo {author} {\bibfnamefont {G.}~\bibnamefont {Peramaiyan}},
  \bibinfo {author} {\bibfnamefont {I.~P.}\ \bibnamefont {Muthuselvam}},
  \bibinfo {author} {\bibfnamefont {C.~J.}\ \bibnamefont {Butler}}, \bibinfo
  {author} {\bibfnamefont {K.}~\bibnamefont {Dimitri}}, \bibinfo {author}
  {\bibfnamefont {M.}~\bibnamefont {Neupane}}, \bibinfo {author} {\bibfnamefont
  {G.~N.}\ \bibnamefont {Rao}}, \bibinfo {author} {\bibfnamefont {M.-T.}\
  \bibnamefont {Lin}}, \ and\ \bibinfo {author} {\bibfnamefont {F.~C.}\
  \bibnamefont {Chou}},\ }\bibfield  {title} {\enquote {\bibinfo {title}
  {Crystal growth of dirac semimetal {ZrSiS} with high magnetoresistance and
  mobility},}\ }\href@noop {} {\bibfield  {journal} {\bibinfo  {journal} {Sci.
  Rep.}\ }\textbf {\bibinfo {volume} {7}},\ \bibinfo {pages} {40603} (\bibinfo
  {year} {2017})}\BibitemShut {NoStop}%
\bibitem [{\citenamefont {Lv}\ \emph {et~al.}(2016)\citenamefont {Lv},
  \citenamefont {Zhang}, \citenamefont {Li}, \citenamefont {Yao}, \citenamefont
  {Chen}, \citenamefont {Zhou}, \citenamefont {Zhang}, \citenamefont {Lu},\
  and\ \citenamefont {Chen}}]{Lv.2016}%
  \BibitemOpen
  \bibfield  {author} {\bibinfo {author} {\bibfnamefont {Y.-Y.}\ \bibnamefont
  {Lv}}, \bibinfo {author} {\bibfnamefont {B.-B.}\ \bibnamefont {Zhang}},
  \bibinfo {author} {\bibfnamefont {X.}~\bibnamefont {Li}}, \bibinfo {author}
  {\bibfnamefont {S.-H.}\ \bibnamefont {Yao}}, \bibinfo {author} {\bibfnamefont
  {Y.~B.}\ \bibnamefont {Chen}}, \bibinfo {author} {\bibfnamefont
  {J.}~\bibnamefont {Zhou}}, \bibinfo {author} {\bibfnamefont {S.-T.}\
  \bibnamefont {Zhang}}, \bibinfo {author} {\bibfnamefont {M.-H.}\ \bibnamefont
  {Lu}}, \ and\ \bibinfo {author} {\bibfnamefont {Y.-F.}\ \bibnamefont
  {Chen}},\ }\bibfield  {title} {\enquote {\bibinfo {title} {Extremely large
  and significantly anisotropic magnetoresistance in {ZrSiS} single
  crystals},}\ }\href@noop {} {\bibfield  {journal} {\bibinfo  {journal} {Appl.
  Phys. Lett.}\ }\textbf {\bibinfo {volume} {108}},\ \bibinfo {pages} {244101}
  (\bibinfo {year} {2016})}\BibitemShut {NoStop}%
\bibitem [{\citenamefont {Singha}\ \emph {et~al.}(2017)\citenamefont {Singha},
  \citenamefont {Pariari}, \citenamefont {Satpati},\ and\ \citenamefont
  {Mandal}}]{Singha.2017}%
  \BibitemOpen
  \bibfield  {author} {\bibinfo {author} {\bibfnamefont {R.}~\bibnamefont
  {Singha}}, \bibinfo {author} {\bibfnamefont {A.~K.}\ \bibnamefont {Pariari}},
  \bibinfo {author} {\bibfnamefont {B.}~\bibnamefont {Satpati}}, \ and\
  \bibinfo {author} {\bibfnamefont {P.}~\bibnamefont {Mandal}},\ }\bibfield
  {title} {\enquote {\bibinfo {title} {Large nonsaturating magnetoresistance
  and signature of nondegenerate dirac nodes in {ZrSiS}},}\ }\href@noop {}
  {\bibfield  {journal} {\bibinfo  {journal} {PNAS}\ }\textbf {\bibinfo
  {volume} {114}},\ \bibinfo {pages} {2468} (\bibinfo {year}
  {2017})}\BibitemShut {NoStop}%
\bibitem [{\citenamefont {Schilling}\ \emph {et~al.}(2017)\citenamefont
  {Schilling}, \citenamefont {Schoop}, \citenamefont {Lotsch}, \citenamefont
  {Dressel},\ and\ \citenamefont {Pronin}}]{Schilling.2017}%
  \BibitemOpen
  \bibfield  {author} {\bibinfo {author} {\bibfnamefont {M.~B.}\ \bibnamefont
  {Schilling}}, \bibinfo {author} {\bibfnamefont {L.~M.}\ \bibnamefont
  {Schoop}}, \bibinfo {author} {\bibfnamefont {B.~V.}\ \bibnamefont {Lotsch}},
  \bibinfo {author} {\bibfnamefont {M.}~\bibnamefont {Dressel}}, \ and\
  \bibinfo {author} {\bibfnamefont {A.~V.}\ \bibnamefont {Pronin}},\ }\bibfield
   {title} {\enquote {\bibinfo {title} {Flat optical conductivity in {ZrSiS}
  due to two-dimensional dirac bands},}\ }\href@noop {} {\bibfield  {journal}
  {\bibinfo  {journal} {Phys. Rev. Lett.}\ }\textbf {\bibinfo {volume} {119}},\
  \bibinfo {pages} {187401} (\bibinfo {year} {2017})}\BibitemShut {NoStop}%
\bibitem [{\citenamefont {Ebad-Allah}\ \emph
  {et~al.}(2019{\natexlab{a}})\citenamefont {Ebad-Allah}, \citenamefont
  {Afonso}, \citenamefont {Krottenm\"uller}, \citenamefont {Hu}, \citenamefont
  {Zhu}, \citenamefont {Mao}, \citenamefont {Kunes},\ and\ \citenamefont
  {Kuntscher}}]{Ebad-Allah.2019}%
  \BibitemOpen
  \bibfield  {author} {\bibinfo {author} {\bibfnamefont {J.}~\bibnamefont
  {Ebad-Allah}}, \bibinfo {author} {\bibfnamefont {J.~F.}\ \bibnamefont
  {Afonso}}, \bibinfo {author} {\bibfnamefont {M.}~\bibnamefont
  {Krottenm\"uller}}, \bibinfo {author} {\bibfnamefont {J.}~\bibnamefont {Hu}},
  \bibinfo {author} {\bibfnamefont {Y.~L.}\ \bibnamefont {Zhu}}, \bibinfo
  {author} {\bibfnamefont {Z.~Q.}\ \bibnamefont {Mao}}, \bibinfo {author}
  {\bibfnamefont {J.}~\bibnamefont {Kunes}}, \ and\ \bibinfo {author}
  {\bibfnamefont {C.~A.}\ \bibnamefont {Kuntscher}},\ }\bibfield  {title}
  {\enquote {\bibinfo {title} {Chemical pressure effect on the optical
  conductivity of the nodal-line semimetals {ZrSi$Y$ ($Y$=S, Se, Te) and
  ZrGe$Y$ ($Y$=S, Te)}},}\ }\href@noop {} {\bibfield  {journal} {\bibinfo
  {journal} {Phys. Rev. B}\ }\textbf {\bibinfo {volume} {99}},\ \bibinfo
  {pages} {125154} (\bibinfo {year} {2019}{\natexlab{a}})}\BibitemShut
  {NoStop}%
\bibitem [{\citenamefont {Perdew}\ \emph {et~al.}(1996)\citenamefont {Perdew},
  \citenamefont {Burke},\ and\ \citenamefont {Ernzerhof}}]{Perdew.1996}%
  \BibitemOpen
  \bibfield  {author} {\bibinfo {author} {\bibfnamefont {J.~P.}\ \bibnamefont
  {Perdew}}, \bibinfo {author} {\bibfnamefont {K.}~\bibnamefont {Burke}}, \
  and\ \bibinfo {author} {\bibfnamefont {M.}~\bibnamefont {Ernzerhof}},\
  }\bibfield  {title} {\enquote {\bibinfo {title} {Generalized gradient
  approximation made simple},}\ }\href {\doibase 10.1103/PhysRevLett.77.3865}
  {\bibfield  {journal} {\bibinfo  {journal} {Phys. Rev. Lett.}\ }\textbf
  {\bibinfo {volume} {77}},\ \bibinfo {pages} {3865} (\bibinfo {year}
  {1996})}\BibitemShut {NoStop}%
\bibitem [{Sup()}]{Suppl}%
  \BibitemOpen
  \href@noop {} {}\bibinfo {note} {{See Supplemental Material at --- for
  details about sample preparation, polarization-dependent reflectivity
  measurements at ambient and high pressure, analysis of reflectivity and
  optical conductivity spectra, XRD measurements under pressure, and
  theoretical calculations, which includes Refs.\
  \cite{Bensch.1995,Mao.1986,Syassen.2008,Eremets.1992,Merrill.1974,boehler_new_2004,Moggach.2008,piermarini_calibration_1975,Dewaele.2008,Kantor,piermarini_hydrostatic_1973,crysalis,Kresse.1993,Kresse.1996,Blochl.1994,Hedin.1965,Hybertsen.1986,Shishkin.2006,Dancoff.1950,Paier.2008}.}}\BibitemShut
  {Stop}%
\bibitem [{\citenamefont {Novak}\ \emph {et~al.}(2019)\citenamefont {Novak},
  \citenamefont {Zhang}, \citenamefont {Orbani\'{c}}, \citenamefont
  {Bili\v{s}kov}, \citenamefont {Eguchi}, \citenamefont {Paschen},
  \citenamefont {Kimura}, \citenamefont {Wang}, \citenamefont {Osada},
  \citenamefont {Uchida}, \citenamefont {Sato}, \citenamefont {Wu},
  \citenamefont {Yazyev},\ and\ \citenamefont {Kokanovi\'{c}}}]{Novak.2019}%
  \BibitemOpen
  \bibfield  {author} {\bibinfo {author} {\bibfnamefont {M.}~\bibnamefont
  {Novak}}, \bibinfo {author} {\bibfnamefont {S.~N.}\ \bibnamefont {Zhang}},
  \bibinfo {author} {\bibfnamefont {F.}~\bibnamefont {Orbani\'{c}}}, \bibinfo
  {author} {\bibfnamefont {N.}~\bibnamefont {Bili\v{s}kov}}, \bibinfo {author}
  {\bibfnamefont {G.}~\bibnamefont {Eguchi}}, \bibinfo {author} {\bibfnamefont
  {S.}~\bibnamefont {Paschen}}, \bibinfo {author} {\bibfnamefont
  {A.}~\bibnamefont {Kimura}}, \bibinfo {author} {\bibfnamefont {X.~X.}\
  \bibnamefont {Wang}}, \bibinfo {author} {\bibfnamefont {T.}~\bibnamefont
  {Osada}}, \bibinfo {author} {\bibfnamefont {K.}~\bibnamefont {Uchida}},
  \bibinfo {author} {\bibfnamefont {M.}~\bibnamefont {Sato}}, \bibinfo {author}
  {\bibfnamefont {Q.~S.}\ \bibnamefont {Wu}}, \bibinfo {author} {\bibfnamefont
  {O.~V.}\ \bibnamefont {Yazyev}}, \ and\ \bibinfo {author} {\bibfnamefont
  {I}~\bibnamefont {Kokanovi\'{c}}},\ }\bibfield  {title} {\enquote {\bibinfo
  {title} {Highly anisotropic interlayer magnetoresitance in {ZrSiS} nodal-line
  dirac semimetal},}\ }\href@noop {} {\bibfield  {journal} {\bibinfo  {journal}
  {Phys. Rev.B}\ }\textbf {\bibinfo {volume} {100}},\ \bibinfo {pages} {085137}
  (\bibinfo {year} {2019})}\BibitemShut {NoStop}%
\bibitem [{\citenamefont {Shirer}\ \emph {et~al.}(2019)\citenamefont {Shirer},
  \citenamefont {Modic}, \citenamefont {Zimmerling}, \citenamefont {Bachmann},
  \citenamefont {König}, \citenamefont {Moll}, \citenamefont {Schoop},\ and\
  \citenamefont {Mackenzie}}]{Shirer.2019}%
  \BibitemOpen
  \bibfield  {author} {\bibinfo {author} {\bibfnamefont {K.~R.}\ \bibnamefont
  {Shirer}}, \bibinfo {author} {\bibfnamefont {K.~A.}\ \bibnamefont {Modic}},
  \bibinfo {author} {\bibfnamefont {T.}~\bibnamefont {Zimmerling}}, \bibinfo
  {author} {\bibfnamefont {M.~D.}\ \bibnamefont {Bachmann}}, \bibinfo {author}
  {\bibfnamefont {M.}~\bibnamefont {König}}, \bibinfo {author} {\bibfnamefont
  {P.~J.~W.}\ \bibnamefont {Moll}}, \bibinfo {author} {\bibfnamefont
  {L.}~\bibnamefont {Schoop}}, \ and\ \bibinfo {author} {\bibfnamefont {A.~P.}\
  \bibnamefont {Mackenzie}},\ }\bibfield  {title} {\enquote {\bibinfo {title}
  {Out-of-plane transport in {ZrSiS} and {ZrSiSe} microstructures},}\
  }\href@noop {} {\bibfield  {journal} {\bibinfo  {journal} {APL Mater.}\
  }\textbf {\bibinfo {volume} {7}},\ \bibinfo {pages} {101116} (\bibinfo {year}
  {2019})}\BibitemShut {NoStop}%
\bibitem [{\citenamefont {Zhou}\ \emph {et~al.}(2019)\citenamefont {Zhou},
  \citenamefont {Rudenko},\ and\ \citenamefont {Yuan}}]{Zhou.2019}%
  \BibitemOpen
  \bibfield  {author} {\bibinfo {author} {\bibfnamefont {W.}~\bibnamefont
  {Zhou}}, \bibinfo {author} {\bibfnamefont {A.~N.}\ \bibnamefont {Rudenko}}, \
  and\ \bibinfo {author} {\bibfnamefont {S.}~\bibnamefont {Yuan}},\ }\bibfield
  {title} {\enquote {\bibinfo {title} {Effect of mechanical strain on the
  optical properties of nodel-line semimetal {ZrSiS}},}\ }\href@noop {}
  {\bibfield  {journal} {\bibinfo  {journal} {Adv. Electon. Mater.}\ }\textbf
  {\bibinfo {volume} {6}},\ \bibinfo {pages} {1900860} (\bibinfo {year}
  {2019})}\BibitemShut {NoStop}%
\bibitem [{\citenamefont {Habe}\ and\ \citenamefont
  {Koshino}(2018)}]{Habe.2018}%
  \BibitemOpen
  \bibfield  {author} {\bibinfo {author} {\bibfnamefont {T.}~\bibnamefont
  {Habe}}\ and\ \bibinfo {author} {\bibfnamefont {M.}~\bibnamefont {Koshino}},\
  }\bibfield  {title} {\enquote {\bibinfo {title} {Dynamical conductivity in
  the topological nodal-line semimetal {ZrSiS}},}\ }\href@noop {} {\bibfield
  {journal} {\bibinfo  {journal} {Phys Rev. B}\ }\textbf {\bibinfo {volume}
  {98}},\ \bibinfo {pages} {125201} (\bibinfo {year} {2018})}\BibitemShut
  {NoStop}%
\bibitem [{1()}]{1}%
  \BibitemOpen
  \href@noop {} {}\bibinfo {note} {Please note that we concentrate here on the
  high-energy range, since the temperature dependence of the low-energy optical
  conductivity for {\bf E}$\parallel $$ab$ has been discussed in detail in
  Ref.\ \cite {Schilling.2017}.}\BibitemShut {Stop}%
\bibitem [{2()}]{2}%
  \BibitemOpen
  \href@noop {} {}\bibinfo {note} {It is interesting to note, that the
  temperature-dependent shift of the L4 peak slightly changes its slope between
  150 and 100~K. In the same temperature range, the temperature dependence of
  the dc resistivity changes \cite {Singha.2017} and several Raman modes show
  an anomaly in their position and width \cite {Singha.2018}, which was
  attributed to an interplay between electron and phonon degrees of
  freedom.}\BibitemShut {Stop}%
\bibitem [{\citenamefont {Young}\ and\ \citenamefont
  {Kane}(2015)}]{Young.2015}%
  \BibitemOpen
  \bibfield  {author} {\bibinfo {author} {\bibfnamefont {S.~M.}\ \bibnamefont
  {Young}}\ and\ \bibinfo {author} {\bibfnamefont {C.~L.}\ \bibnamefont
  {Kane}},\ }\bibfield  {title} {\enquote {\bibinfo {title} {Dirac semimetals
  in two dimensions},}\ }\href@noop {} {\bibfield  {journal} {\bibinfo
  {journal} {Phys. Rev. Lett.}\ }\textbf {\bibinfo {volume} {115}},\ \bibinfo
  {pages} {126803} (\bibinfo {year} {2015})}\BibitemShut {NoStop}%
\bibitem [{3()}]{3}%
  \BibitemOpen
  \href@noop {} {}\bibinfo {note} {We cannot completely rule out that the F2
  peak already exists at ambient pressure, but cannot be resolved due to small
  oscillator strength and overlap with higher-energy interband
  transitions.}\BibitemShut {Stop}%
\bibitem [{\citenamefont {Gu}\ \emph {et~al.}(2019)\citenamefont {Gu},
  \citenamefont {Hu}, \citenamefont {Chen}, \citenamefont {Guo}, \citenamefont
  {Fu}, \citenamefont {Zhou}, \citenamefont {An}, \citenamefont {Zhou},
  \citenamefont {Zhang}, \citenamefont {Xi}, \citenamefont {Gu}, \citenamefont
  {Park}, \citenamefont {Shu}, \citenamefont {Yang}, \citenamefont {Pi},
  \citenamefont {Zhang}, \citenamefont {Yao}, \citenamefont {Yang},
  \citenamefont {Zhou}, \citenamefont {Sun}, \citenamefont {Mao},\ and\
  \citenamefont {Tian}}]{Gu.2019}%
  \BibitemOpen
  \bibfield  {author} {\bibinfo {author} {\bibfnamefont {C.~C.}\ \bibnamefont
  {Gu}}, \bibinfo {author} {\bibfnamefont {J.}~\bibnamefont {Hu}}, \bibinfo
  {author} {\bibfnamefont {X.~L.}\ \bibnamefont {Chen}}, \bibinfo {author}
  {\bibfnamefont {Z.~P.}\ \bibnamefont {Guo}}, \bibinfo {author} {\bibfnamefont
  {B.~T.}\ \bibnamefont {Fu}}, \bibinfo {author} {\bibfnamefont {Y.~H.}\
  \bibnamefont {Zhou}}, \bibinfo {author} {\bibfnamefont {C.}~\bibnamefont
  {An}}, \bibinfo {author} {\bibfnamefont {Y.}~\bibnamefont {Zhou}}, \bibinfo
  {author} {\bibfnamefont {R.~R.}\ \bibnamefont {Zhang}}, \bibinfo {author}
  {\bibfnamefont {C.~Y.}\ \bibnamefont {Xi}}, \bibinfo {author} {\bibfnamefont
  {Q.~Y.}\ \bibnamefont {Gu}}, \bibinfo {author} {\bibfnamefont
  {C.}~\bibnamefont {Park}}, \bibinfo {author} {\bibfnamefont {H.~Y.}\
  \bibnamefont {Shu}}, \bibinfo {author} {\bibfnamefont {W.~G.}\ \bibnamefont
  {Yang}}, \bibinfo {author} {\bibfnamefont {L.}~\bibnamefont {Pi}}, \bibinfo
  {author} {\bibfnamefont {Y.~H.}\ \bibnamefont {Zhang}}, \bibinfo {author}
  {\bibfnamefont {Y.~G.}\ \bibnamefont {Yao}}, \bibinfo {author} {\bibfnamefont
  {Z.~R.}\ \bibnamefont {Yang}}, \bibinfo {author} {\bibfnamefont {J.~H.}\
  \bibnamefont {Zhou}}, \bibinfo {author} {\bibfnamefont {J.}~\bibnamefont
  {Sun}}, \bibinfo {author} {\bibfnamefont {Z.~Q.}\ \bibnamefont {Mao}}, \ and\
  \bibinfo {author} {\bibfnamefont {M.~L.}\ \bibnamefont {Tian}},\ }\bibfield
  {title} {\enquote {\bibinfo {title} {Experimental evidence of crystal
  symmetry protection for the topological nodal line semimetal state in
  {ZrSiS}},}\ }\href@noop {} {\bibfield  {journal} {\bibinfo  {journal} {Phys.
  Rev. B}\ }\textbf {\bibinfo {volume} {100}},\ \bibinfo {pages} {205124}
  (\bibinfo {year} {2019})}\BibitemShut {NoStop}%
\bibitem [{\citenamefont {VanGennep}\ \emph {et~al.}(2019)\citenamefont
  {VanGennep}, \citenamefont {Paul}, \citenamefont {Yerger}, \citenamefont
  {Weir}, \citenamefont {Vohra},\ and\ \citenamefont
  {Hamlin}}]{VanGennep.2019}%
  \BibitemOpen
  \bibfield  {author} {\bibinfo {author} {\bibfnamefont {D.}~\bibnamefont
  {VanGennep}}, \bibinfo {author} {\bibfnamefont {T.~A.}\ \bibnamefont {Paul}},
  \bibinfo {author} {\bibfnamefont {C.~W.}\ \bibnamefont {Yerger}}, \bibinfo
  {author} {\bibfnamefont {S.~T.}\ \bibnamefont {Weir}}, \bibinfo {author}
  {\bibfnamefont {Y.~K.}\ \bibnamefont {Vohra}}, \ and\ \bibinfo {author}
  {\bibfnamefont {J.~J.}\ \bibnamefont {Hamlin}},\ }\bibfield  {title}
  {\enquote {\bibinfo {title} {Possible pressure-induced topological quantum
  phase transition in the nodal line semimetal {ZrSiS}},}\ }\href@noop {}
  {\bibfield  {journal} {\bibinfo  {journal} {Phys. Rev. B}\ }\textbf {\bibinfo
  {volume} {99}},\ \bibinfo {pages} {085204} (\bibinfo {year}
  {2019})}\BibitemShut {NoStop}%
\bibitem [{\citenamefont {Singha}\ \emph {et~al.}(2018)\citenamefont {Singha},
  \citenamefont {Samanta}, \citenamefont {Chatterjee}, \citenamefont {Pariari},
  \citenamefont {Majumdar}, \citenamefont {Satpati}, \citenamefont {Wang},
  \citenamefont {Singha},\ and\ \citenamefont {Mandal}}]{Singha.2018}%
  \BibitemOpen
  \bibfield  {author} {\bibinfo {author} {\bibfnamefont {R.}~\bibnamefont
  {Singha}}, \bibinfo {author} {\bibfnamefont {S.}~\bibnamefont {Samanta}},
  \bibinfo {author} {\bibfnamefont {S.}~\bibnamefont {Chatterjee}}, \bibinfo
  {author} {\bibfnamefont {A.}~\bibnamefont {Pariari}}, \bibinfo {author}
  {\bibfnamefont {D.}~\bibnamefont {Majumdar}}, \bibinfo {author}
  {\bibfnamefont {B.}~\bibnamefont {Satpati}}, \bibinfo {author} {\bibfnamefont
  {L.}~\bibnamefont {Wang}}, \bibinfo {author} {\bibfnamefont {A.}~\bibnamefont
  {Singha}}, \ and\ \bibinfo {author} {\bibfnamefont {P.}~\bibnamefont
  {Mandal}},\ }\bibfield  {title} {\enquote {\bibinfo {title} {Probing lattice
  dynamics and electron-phonon coupling in the topological nodal-line semimetal
  {ZrSiS}},}\ }\href@noop {} {\bibfield  {journal} {\bibinfo  {journal} {Phys
  Rev. B}\ }\textbf {\bibinfo {volume} {97}},\ \bibinfo {pages} {094112}
  (\bibinfo {year} {2018})}\BibitemShut {NoStop}%
\bibitem [{\citenamefont {Murnaghan}(1944)}]{Murnaghan.1944}%
  \BibitemOpen
  \bibfield  {author} {\bibinfo {author} {\bibfnamefont {F.~D.}\ \bibnamefont
  {Murnaghan}},\ }\bibfield  {title} {\enquote {\bibinfo {title} {The
  compressibility of media under extreme pressures},}\ }\href@noop {}
  {\bibfield  {journal} {\bibinfo  {journal} {Proc. Natl. Acad. Sci.}\ }\textbf
  {\bibinfo {volume} {30}},\ \bibinfo {pages} {244} (\bibinfo {year}
  {1944})}\BibitemShut {NoStop}%
\bibitem [{\citenamefont {Salmankurt}\ and\ \citenamefont
  {Duman}(2016)}]{Salmankurt.2016}%
  \BibitemOpen
  \bibfield  {author} {\bibinfo {author} {\bibfnamefont {B.}~\bibnamefont
  {Salmankurt}}\ and\ \bibinfo {author} {\bibfnamefont {S.}~\bibnamefont
  {Duman}},\ }\bibfield  {title} {\enquote {\bibinfo {title} {First-principles
  study of structural, mechanical, lattice dynamical and thermal properties of
  nodal-line semimetals {ZrXY (X=Si,Ge; Y=S,Se)}},}\ }\href@noop {} {\bibfield
  {journal} {\bibinfo  {journal} {Phil. Mag.}\ }\textbf {\bibinfo {volume}
  {97}},\ \bibinfo {pages} {175} (\bibinfo {year} {2016})}\BibitemShut
  {NoStop}%
\bibitem [{\citenamefont {Kirby}\ \emph {et~al.}(2020)\citenamefont {Kirby},
  \citenamefont {Ferrenti}, \citenamefont {Weinberg}, \citenamefont {Klemenz},
  \citenamefont {Oudah}, \citenamefont {Lei}, \citenamefont {Weber},
  \citenamefont {Fausti}, \citenamefont {Scholes},\ and\ \citenamefont
  {Schoop}}]{Kirby.2020}%
  \BibitemOpen
  \bibfield  {author} {\bibinfo {author} {\bibfnamefont {R.~J.}\ \bibnamefont
  {Kirby}}, \bibinfo {author} {\bibfnamefont {A.}~\bibnamefont {Ferrenti}},
  \bibinfo {author} {\bibfnamefont {C.}~\bibnamefont {Weinberg}}, \bibinfo
  {author} {\bibfnamefont {S.}~\bibnamefont {Klemenz}}, \bibinfo {author}
  {\bibfnamefont {M.}~\bibnamefont {Oudah}}, \bibinfo {author} {\bibfnamefont
  {S.}~\bibnamefont {Lei}}, \bibinfo {author} {\bibfnamefont {C.~P.}\
  \bibnamefont {Weber}}, \bibinfo {author} {\bibfnamefont {D.}~\bibnamefont
  {Fausti}}, \bibinfo {author} {\bibfnamefont {G.~D.}\ \bibnamefont {Scholes}},
  \ and\ \bibinfo {author} {\bibfnamefont {L.~M.}\ \bibnamefont {Schoop}},\
  }\bibfield  {title} {\enquote {\bibinfo {title} {Transient drude response
  dominates near-infrared pump-probe reflectivity in nodal-line semimetals
  {ZrSiS} and {ZrSiSe}},}\ }\href@noop {} {\bibfield  {journal} {\bibinfo
  {journal} {J. Phys. Chem. Lett.}\ }\textbf {\bibinfo {volume} {11}},\
  \bibinfo {pages} {6105} (\bibinfo {year} {2020})}\BibitemShut {NoStop}%
\bibitem [{\citenamefont {Ebad-Allah}\ \emph
  {et~al.}(2019{\natexlab{b}})\citenamefont {Ebad-Allah}, \citenamefont
  {Krottenm\"uller}, \citenamefont {Hu}, \citenamefont {Zhu}, \citenamefont
  {Mao},\ and\ \citenamefont {Kuntscher}}]{Ebad-Allah.2019a}%
  \BibitemOpen
  \bibfield  {author} {\bibinfo {author} {\bibfnamefont {J.}~\bibnamefont
  {Ebad-Allah}}, \bibinfo {author} {\bibfnamefont {M.}~\bibnamefont
  {Krottenm\"uller}}, \bibinfo {author} {\bibfnamefont {J.}~\bibnamefont {Hu}},
  \bibinfo {author} {\bibfnamefont {Y.~L.}\ \bibnamefont {Zhu}}, \bibinfo
  {author} {\bibfnamefont {Z.~Q.}\ \bibnamefont {Mao}}, \ and\ \bibinfo
  {author} {\bibfnamefont {C.~A.}\ \bibnamefont {Kuntscher}},\ }\bibfield
  {title} {\enquote {\bibinfo {title} {Infrared spectroscopy study of the
  nodal-line semimetal candidate zrsite under pressure: Hints for
  pressure-induced phase transitions},}\ }\href@noop {} {\bibfield  {journal}
  {\bibinfo  {journal} {Phys. Rev. B}\ }\textbf {\bibinfo {volume} {99}},\
  \bibinfo {pages} {245133} (\bibinfo {year} {2019}{\natexlab{b}})}\BibitemShut
  {NoStop}%
\bibitem [{\citenamefont {Krottenm\"uller}\ \emph {et~al.}(2020)\citenamefont
  {Krottenm\"uller}, \citenamefont {V\"ost}, \citenamefont {Unglert},
  \citenamefont {Ebad-Allah}, \citenamefont {Eirckerling}, \citenamefont
  {Volkmer}, \citenamefont {Hu}, \citenamefont {Zhu}, \citenamefont {Mao},
  \citenamefont {Scherer},\ and\ \citenamefont
  {Kuntscher}}]{Krottenmuller.2020}%
  \BibitemOpen
  \bibfield  {author} {\bibinfo {author} {\bibfnamefont {M.}~\bibnamefont
  {Krottenm\"uller}}, \bibinfo {author} {\bibfnamefont {M.}~\bibnamefont
  {V\"ost}}, \bibinfo {author} {\bibfnamefont {N.}~\bibnamefont {Unglert}},
  \bibinfo {author} {\bibfnamefont {J.}~\bibnamefont {Ebad-Allah}}, \bibinfo
  {author} {\bibfnamefont {G.}~\bibnamefont {Eirckerling}}, \bibinfo {author}
  {\bibfnamefont {D.}~\bibnamefont {Volkmer}}, \bibinfo {author} {\bibfnamefont
  {J.}~\bibnamefont {Hu}}, \bibinfo {author} {\bibfnamefont {Y.~L.}\
  \bibnamefont {Zhu}}, \bibinfo {author} {\bibfnamefont {Z.~Q.}\ \bibnamefont
  {Mao}}, \bibinfo {author} {\bibfnamefont {W.}~\bibnamefont {Scherer}}, \ and\
  \bibinfo {author} {\bibfnamefont {C.~A.}\ \bibnamefont {Kuntscher}},\
  }\bibfield  {title} {\enquote {\bibinfo {title} {Indications for lifshitz
  transitions in the nodal-line semimetal zrsite induced by interlayer
  interaction},}\ }\href@noop {} {\bibfield  {journal} {\bibinfo  {journal}
  {Phys. Rev. B.}\ }\textbf {\bibinfo {volume} {101}},\ \bibinfo {pages}
  {081108(R)} (\bibinfo {year} {2020})}\BibitemShut {NoStop}%
\bibitem [{\citenamefont {Hebel}\ and\ \citenamefont
  {Slichter}(1959)}]{Hebel1959}%
  \BibitemOpen
  \bibfield  {author} {\bibinfo {author} {\bibfnamefont {L.~C.}\ \bibnamefont
  {Hebel}}\ and\ \bibinfo {author} {\bibfnamefont {C.~P.}\ \bibnamefont
  {Slichter}},\ }\bibfield  {title} {\enquote {\bibinfo {title} {Nuclear spin
  relaxation in normal and superconducting aluminum},}\ }\href {\doibase
  10.1103/PhysRev.113.1504} {\bibfield  {journal} {\bibinfo  {journal} {Phys.
  Rev.}\ }\textbf {\bibinfo {volume} {113}},\ \bibinfo {pages} {1504} (\bibinfo
  {year} {1959})}\BibitemShut {NoStop}%
\bibitem [{4()}]{4}%
  \BibitemOpen
  \href@noop {} {}\bibinfo {note} {We note that the F1 peak derives its
  spectral weight from the high-energy rather than low-energy region. This is
  opposite to the excitonic insulator scenario, where the spectral weight comes
  from the Drude peak or low-energy region in general.}\BibitemShut {Stop}%
\bibitem [{\citenamefont {Samaneh~Ataei}\ \emph {et~al.}(2021)\citenamefont
  {Samaneh~Ataei}, \citenamefont {Varsano}, \citenamefont {Molinari},\ and\
  \citenamefont {Rontani}}]{Samaneh.2021}%
  \BibitemOpen
  \bibfield  {author} {\bibinfo {author} {\bibfnamefont {S.}~\bibnamefont
  {Samaneh~Ataei}}, \bibinfo {author} {\bibfnamefont {D.}~\bibnamefont
  {Varsano}}, \bibinfo {author} {\bibfnamefont {E.}~\bibnamefont {Molinari}}, \
  and\ \bibinfo {author} {\bibfnamefont {M.}~\bibnamefont {Rontani}},\
  }\bibfield  {title} {\enquote {\bibinfo {title} {Evidence of ideal excitonic
  insulator in bulk {MoS$_2$} under pressure},}\ }\href@noop {} {\bibfield
  {journal} {\bibinfo  {journal} {PNAS}\ }\textbf {\bibinfo {volume} {118}},\
  \bibinfo {pages} {e2010110118} (\bibinfo {year} {2021})}\BibitemShut
  {NoStop}%
\bibitem [{\citenamefont {Bensch}\ \emph {et~al.}(1995)\citenamefont {Bensch},
  \citenamefont {Helmer}, \citenamefont {Muhler}, \citenamefont {Ebert},\ and\
  \citenamefont {Knecht}}]{Bensch.1995}%
  \BibitemOpen
  \bibfield  {author} {\bibinfo {author} {\bibfnamefont {W.}~\bibnamefont
  {Bensch}}, \bibinfo {author} {\bibfnamefont {O.}~\bibnamefont {Helmer}},
  \bibinfo {author} {\bibfnamefont {M.}~\bibnamefont {Muhler}}, \bibinfo
  {author} {\bibfnamefont {H.}~\bibnamefont {Ebert}}, \ and\ \bibinfo {author}
  {\bibfnamefont {M.}~\bibnamefont {Knecht}},\ }\bibfield  {title} {\enquote
  {\bibinfo {title} {Experimental and theoretical bandstructure of the layer
  compound {ZrSiTe}},}\ }\href@noop {} {\bibfield  {journal} {\bibinfo
  {journal} {J. Phys. Chem.}\ }\textbf {\bibinfo {volume} {99}},\ \bibinfo
  {pages} {3326} (\bibinfo {year} {1995})}\BibitemShut {NoStop}%
\bibitem [{\citenamefont {Mao}\ \emph {et~al.}(1986)\citenamefont {Mao},
  \citenamefont {Xu},\ and\ \citenamefont {Bell}}]{Mao.1986}%
  \BibitemOpen
  \bibfield  {author} {\bibinfo {author} {\bibfnamefont {H.~K.}\ \bibnamefont
  {Mao}}, \bibinfo {author} {\bibfnamefont {J.}~\bibnamefont {Xu}}, \ and\
  \bibinfo {author} {\bibfnamefont {P.~M.}\ \bibnamefont {Bell}},\ }\bibfield
  {title} {\enquote {\bibinfo {title} {Calibration of the ruby pressure gauge
  to 800 kbar under quasi-hydrostatic conditions},}\ }\href@noop {} {\bibfield
  {journal} {\bibinfo  {journal} {J. Geophys. Res.}\ }\textbf {\bibinfo
  {volume} {91}},\ \bibinfo {pages} {4673} (\bibinfo {year}
  {1986})}\BibitemShut {NoStop}%
\bibitem [{\citenamefont {Syassen}(2008)}]{Syassen.2008}%
  \BibitemOpen
  \bibfield  {author} {\bibinfo {author} {\bibfnamefont {K.}~\bibnamefont
  {Syassen}},\ }\bibfield  {title} {\enquote {\bibinfo {title} {Ruby under
  pressure},}\ }\href@noop {} {\bibfield  {journal} {\bibinfo  {journal} {High
  Pressure Res.}\ }\textbf {\bibinfo {volume} {28}},\ \bibinfo {pages} {75}
  (\bibinfo {year} {2008})}\BibitemShut {NoStop}%
\bibitem [{\citenamefont {Eremets}\ and\ \citenamefont
  {Timofeev}(1992)}]{Eremets.1992}%
  \BibitemOpen
  \bibfield  {author} {\bibinfo {author} {\bibfnamefont {M.~I.}\ \bibnamefont
  {Eremets}}\ and\ \bibinfo {author} {\bibfnamefont {Y.~A.}\ \bibnamefont
  {Timofeev}},\ }\href@noop {} {\bibfield  {journal} {\bibinfo  {journal} {Rev.
  Sci. Instrum.}\ }\textbf {\bibinfo {volume} {63}},\ \bibinfo {pages} {3123}
  (\bibinfo {year} {1992})}\BibitemShut {NoStop}%
\bibitem [{\citenamefont {Merrill}\ and\ \citenamefont
  {Bassett}(1974)}]{Merrill.1974}%
  \BibitemOpen
  \bibfield  {author} {\bibinfo {author} {\bibfnamefont {Leo}\ \bibnamefont
  {Merrill}}\ and\ \bibinfo {author} {\bibfnamefont {William~A.}\ \bibnamefont
  {Bassett}},\ }\bibfield  {title} {\enquote {\bibinfo {title} {Miniature
  diamond anvil pressure cell for single crystal x-ray diffraction studies},}\
  }\href {\doibase 10.1063/1.1686607} {\bibfield  {journal} {\bibinfo
  {journal} {Rev. Sci. Instr.}\ }\textbf {\bibinfo {volume} {45}},\ \bibinfo
  {pages} {290} (\bibinfo {year} {1974})}\BibitemShut {NoStop}%
\bibitem [{\citenamefont {Boehler}\ and\ \citenamefont
  {De~Hantsetters}(2004)}]{boehler_new_2004}%
  \BibitemOpen
  \bibfield  {author} {\bibinfo {author} {\bibfnamefont {R.}~\bibnamefont
  {Boehler}}\ and\ \bibinfo {author} {\bibfnamefont {K.}~\bibnamefont
  {De~Hantsetters}},\ }\bibfield  {title} {\enquote {\bibinfo {title} {New
  anvil designs in diamond-cells},}\ }\href@noop {} {\bibfield  {journal}
  {\bibinfo  {journal} {High Pressure Res.}\ }\textbf {\bibinfo {volume}
  {24}},\ \bibinfo {pages} {391} (\bibinfo {year} {2004})}\BibitemShut
  {NoStop}%
\bibitem [{\citenamefont {Moggach}\ \emph {et~al.}(2008)\citenamefont
  {Moggach}, \citenamefont {Allan}, \citenamefont {Parsons},\ and\
  \citenamefont {Warren}}]{Moggach.2008}%
  \BibitemOpen
  \bibfield  {author} {\bibinfo {author} {\bibfnamefont {Stephen~A.}\
  \bibnamefont {Moggach}}, \bibinfo {author} {\bibfnamefont {David~R.}\
  \bibnamefont {Allan}}, \bibinfo {author} {\bibfnamefont {Simon}\ \bibnamefont
  {Parsons}}, \ and\ \bibinfo {author} {\bibfnamefont {John~E.}\ \bibnamefont
  {Warren}},\ }\bibfield  {title} {\enquote {\bibinfo {title} {Incorporation of
  a new design of backing seat and anvil in a {Merrill}–{Bassett} diamond
  anvil cell},}\ }\href {\doibase 10.1107/S0021889808000514} {\bibfield
  {journal} {\bibinfo  {journal} {J. Appl. Cryst.}\ }\textbf {\bibinfo {volume}
  {41}},\ \bibinfo {pages} {249} (\bibinfo {year} {2008})}\BibitemShut
  {NoStop}%
\bibitem [{\citenamefont {Piermarini}\ \emph {et~al.}(1975)\citenamefont
  {Piermarini}, \citenamefont {Block}, \citenamefont {Barnett},\ and\
  \citenamefont {Forman}}]{piermarini_calibration_1975}%
  \BibitemOpen
  \bibfield  {author} {\bibinfo {author} {\bibfnamefont {G.~J.}\ \bibnamefont
  {Piermarini}}, \bibinfo {author} {\bibfnamefont {S.}~\bibnamefont {Block}},
  \bibinfo {author} {\bibfnamefont {J.~D.}\ \bibnamefont {Barnett}}, \ and\
  \bibinfo {author} {\bibfnamefont {R.~A.}\ \bibnamefont {Forman}},\ }\bibfield
   {title} {\enquote {\bibinfo {title} {Calibration of the pressure dependence
  of the {R1} ruby fluorescence line to 195 kbar},}\ }\href@noop {} {\bibfield
  {journal} {\bibinfo  {journal} {J. Appl. Phys.}\ }\textbf {\bibinfo {volume}
  {46}},\ \bibinfo {pages} {2774} (\bibinfo {year} {1975})}\BibitemShut
  {NoStop}%
\bibitem [{\citenamefont {Dewaele}\ \emph {et~al.}(2008)\citenamefont
  {Dewaele}, \citenamefont {Torrent}, \citenamefont {Loubeyre},\ and\
  \citenamefont {Mezouar}}]{Dewaele.2008}%
  \BibitemOpen
  \bibfield  {author} {\bibinfo {author} {\bibfnamefont {Agn\`es}\ \bibnamefont
  {Dewaele}}, \bibinfo {author} {\bibfnamefont {Marc}\ \bibnamefont {Torrent}},
  \bibinfo {author} {\bibfnamefont {Paul}\ \bibnamefont {Loubeyre}}, \ and\
  \bibinfo {author} {\bibfnamefont {Mohamed}\ \bibnamefont {Mezouar}},\
  }\bibfield  {title} {\enquote {\bibinfo {title} {Compression curves of
  transition metals in the mbar range: Experiments and projector augmented-wave
  calculations},}\ }\href {\doibase 10.1103/PhysRevB.78.104102} {\bibfield
  {journal} {\bibinfo  {journal} {Phys. Rev. B}\ }\textbf {\bibinfo {volume}
  {78}},\ \bibinfo {pages} {104102} (\bibinfo {year} {2008})}\BibitemShut
  {NoStop}%
\bibitem [{\citenamefont {{Innokenty Kantor}}()}]{Kantor}%
  \BibitemOpen
  \bibfield  {author} {\bibinfo {author} {\bibnamefont {{Innokenty Kantor}}},\
  }\href {https://millenia.cars.aps.anl.gov/gsecars/ruby/ruby.htm} {\enquote
  {\bibinfo {title} {Fluorescense pressure calculation and thermocouple
  tools},}\ }\bibinfo {note}
  {[https://millenia.cars.aps.anl.gov/gsecars/ruby/ruby.htm; accessed
  13-August-2020]}\BibitemShut {NoStop}%
\bibitem [{\citenamefont {Piermarini}\ \emph {et~al.}(1973)\citenamefont
  {Piermarini}, \citenamefont {Block},\ and\ \citenamefont
  {Barnett}}]{piermarini_hydrostatic_1973}%
  \BibitemOpen
  \bibfield  {author} {\bibinfo {author} {\bibfnamefont {G.~J.}\ \bibnamefont
  {Piermarini}}, \bibinfo {author} {\bibfnamefont {S.}~\bibnamefont {Block}}, \
  and\ \bibinfo {author} {\bibfnamefont {J.D.}\ \bibnamefont {Barnett}},\
  }\bibfield  {title} {\enquote {\bibinfo {title} {Hydrostatic limits in
  liquids and solids to 100 kbar},}\ }\href@noop {} {\bibfield  {journal}
  {\bibinfo  {journal} {J. Appl. Phys.}\ }\textbf {\bibinfo {volume} {44}},\
  \bibinfo {pages} {5377} (\bibinfo {year} {1973})}\BibitemShut {NoStop}%
\bibitem [{\citenamefont {OD}(2017)}]{crysalis}%
  \BibitemOpen
  \bibfield  {author} {\bibinfo {author} {\bibfnamefont {Rigaku}\ \bibnamefont
  {OD}},\ }\bibfield  {title} {\enquote {\bibinfo {title} {{CrysAlisPro
  1.171.38.46}},}\ }\href@noop {} {\  (\bibinfo {year} {2017})}\BibitemShut
  {NoStop}%
\bibitem [{\citenamefont {Kresse}\ and\ \citenamefont
  {Hafner}(1993)}]{Kresse.1993}%
  \BibitemOpen
  \bibfield  {author} {\bibinfo {author} {\bibfnamefont {G.}~\bibnamefont
  {Kresse}}\ and\ \bibinfo {author} {\bibfnamefont {J.}~\bibnamefont
  {Hafner}},\ }\bibfield  {title} {\enquote {\bibinfo {title} {Ab initio
  molecular dynamics for liquid metals},}\ }\href {\doibase
  10.1103/PhysRevB.47.558} {\bibfield  {journal} {\bibinfo  {journal} {Phys.
  Rev. B}\ }\textbf {\bibinfo {volume} {47}},\ \bibinfo {pages} {558} (\bibinfo
  {year} {1993})}\BibitemShut {NoStop}%
\bibitem [{\citenamefont {Kresse}\ and\ \citenamefont
  {Furthm\"uller}(1996)}]{Kresse.1996}%
  \BibitemOpen
  \bibfield  {author} {\bibinfo {author} {\bibfnamefont {G.}~\bibnamefont
  {Kresse}}\ and\ \bibinfo {author} {\bibfnamefont {J.}~\bibnamefont
  {Furthm\"uller}},\ }\bibfield  {title} {\enquote {\bibinfo {title} {Efficient
  iterative schemes for ab initio total-energy calculations using a plane-wave
  basis set},}\ }\href {\doibase 10.1103/PhysRevB.54.11169} {\bibfield
  {journal} {\bibinfo  {journal} {Phys. Rev. B}\ }\textbf {\bibinfo {volume}
  {54}},\ \bibinfo {pages} {11169} (\bibinfo {year} {1996})}\BibitemShut
  {NoStop}%
\bibitem [{\citenamefont {Bl\"ochl}(1994)}]{Blochl.1994}%
  \BibitemOpen
  \bibfield  {author} {\bibinfo {author} {\bibfnamefont {P.~E.}\ \bibnamefont
  {Bl\"ochl}},\ }\bibfield  {title} {\enquote {\bibinfo {title} {Projector
  augmented-wave method},}\ }\href {\doibase 10.1103/PhysRevB.50.17953}
  {\bibfield  {journal} {\bibinfo  {journal} {Phys. Rev. B}\ }\textbf {\bibinfo
  {volume} {50}},\ \bibinfo {pages} {17953} (\bibinfo {year}
  {1994})}\BibitemShut {NoStop}%
\bibitem [{\citenamefont {Hedin}(1965)}]{Hedin.1965}%
  \BibitemOpen
  \bibfield  {author} {\bibinfo {author} {\bibfnamefont {Lars}\ \bibnamefont
  {Hedin}},\ }\bibfield  {title} {\enquote {\bibinfo {title} {New method for
  calculating the one-particle {G}reen's {F}unction with application to the
  electron-gas problem},}\ }\href {\doibase 10.1103/PhysRev.139.A796}
  {\bibfield  {journal} {\bibinfo  {journal} {Phys. Rev.}\ }\textbf {\bibinfo
  {volume} {139}},\ \bibinfo {pages} {A796} (\bibinfo {year}
  {1965})}\BibitemShut {NoStop}%
\bibitem [{\citenamefont {Hybertsen}\ and\ \citenamefont
  {Louie}(1986)}]{Hybertsen.1986}%
  \BibitemOpen
  \bibfield  {author} {\bibinfo {author} {\bibfnamefont {Mark~S.}\ \bibnamefont
  {Hybertsen}}\ and\ \bibinfo {author} {\bibfnamefont {Steven~G.}\ \bibnamefont
  {Louie}},\ }\bibfield  {title} {\enquote {\bibinfo {title} {Electron
  correlation in semiconductors and insulators: Band gaps and quasiparticle
  energies},}\ }\href {\doibase 10.1103/PhysRevB.34.5390} {\bibfield  {journal}
  {\bibinfo  {journal} {Phys. Rev. B}\ }\textbf {\bibinfo {volume} {34}},\
  \bibinfo {pages} {5390} (\bibinfo {year} {1986})}\BibitemShut {NoStop}%
\bibitem [{\citenamefont {Shishkin}\ and\ \citenamefont
  {Kresse}(2006)}]{Shishkin.2006}%
  \BibitemOpen
  \bibfield  {author} {\bibinfo {author} {\bibfnamefont {M.}~\bibnamefont
  {Shishkin}}\ and\ \bibinfo {author} {\bibfnamefont {G.}~\bibnamefont
  {Kresse}},\ }\bibfield  {title} {\enquote {\bibinfo {title} {Implementation
  and performance of the frequency-dependent {GW} method within the {PAW}
  framework},}\ }\href {\doibase 10.1103/PhysRevB.74.035101} {\bibfield
  {journal} {\bibinfo  {journal} {Phys. Rev. B}\ }\textbf {\bibinfo {volume}
  {74}},\ \bibinfo {pages} {035101} (\bibinfo {year} {2006})}\BibitemShut
  {NoStop}%
\bibitem [{\citenamefont {Dancoff}(1950)}]{Dancoff.1950}%
  \BibitemOpen
  \bibfield  {author} {\bibinfo {author} {\bibfnamefont {S.~M.}\ \bibnamefont
  {Dancoff}},\ }\bibfield  {title} {\enquote {\bibinfo {title} {Non-adiabatic
  meson theory of nuclear forces},}\ }\href {\doibase 10.1103/PhysRev.78.382}
  {\bibfield  {journal} {\bibinfo  {journal} {Phys. Rev.}\ }\textbf {\bibinfo
  {volume} {78}},\ \bibinfo {pages} {382} (\bibinfo {year} {1950})}\BibitemShut
  {NoStop}%
\bibitem [{\citenamefont {Paier}\ \emph {et~al.}(2008)\citenamefont {Paier},
  \citenamefont {Marsman},\ and\ \citenamefont {Kresse}}]{Paier.2008}%
  \BibitemOpen
  \bibfield  {author} {\bibinfo {author} {\bibfnamefont {Joachim}\ \bibnamefont
  {Paier}}, \bibinfo {author} {\bibfnamefont {Martijn}\ \bibnamefont
  {Marsman}}, \ and\ \bibinfo {author} {\bibfnamefont {Georg}\ \bibnamefont
  {Kresse}},\ }\bibfield  {title} {\enquote {\bibinfo {title} {Dielectric
  properties and excitons for extended systems from hybrid functionals},}\
  }\href {\doibase 10.1103/PhysRevB.78.121201} {\bibfield  {journal} {\bibinfo
  {journal} {Phys. Rev. B}\ }\textbf {\bibinfo {volume} {78}},\ \bibinfo
  {pages} {121201} (\bibinfo {year} {2008})}\BibitemShut {NoStop}%
\end{thebibliography}
\end{document}